# Single-shot phase retrieval in high-energy X-ray grating interferometry

Zhili Wang*, and Dalin Liu

*School of Electronic Science & Applied Physics, Hefei University of Technology, Hefei 230009, China*
*Corresponding author: dywangzl@hfut.edu.cn*

In X-ray phase contrast imaging, phase retrieval from intensity measurements is the key for further quantitative analysis and tomographic reconstructions. In this letter, we present a single-shot approach for quantitative phase retrieval in high-energy X-ray grating interferometry. The proposed approach makes use of the phase-attenuation duality of soft tissues when being imaged with high-energy X-rays. The phase retrieval formula is derived and presented, and tested by numerical experiments including photon shot noise. The good agreement between retrieval results and theoretical values confirms the feasibility of the presented approach.
© 2016 Optical Society of America

---

X-ray grating interferometry has attracted increasing attention recently owing to its high sensitivity [1], its quantitativeness [2,3] and multi-contrast capabilities [4]. Particularly, X-ray grating interferometry can be efficiently implemented with conventional X-ray sources to exploit phase contrast [5,6], thus potentially interesting for a variety of applications including preclinical and clinical imaging [7,8], biomedical imaging [9], materials science [4] etc. Recent studies have seen the development of high-energy X-ray grating interferometry, including Talbot interferometry at photon energies up to 183 keV using synchrotron radiation [2,10,11] and Talbot-Lau interferometry at X-ray energies up to 100 keV with laboratory X-ray sources [12]. As already demonstrated, phase imaging could theoretically be performed at slightly higher X-ray energies compared with conventional absorption imaging, resulting in a substantial reduced dose deposition to the sample [9,13]. This is a fundamental advantage of phase-contrast imaging compared with absorption-based techniques and of crucial importance when medical or even screening applications are envisaged [12].

In X-ray grating interferometry, the image contrast in the recorded radiographs can arise from a mixture of the absorption and phase shift of the imaged object. Therefore, increasing attention has been paid to quantitative phase retrieval from intensity measurements [14-19], which is necessary for quantitative tissue characterization, differentiating healthy and diseased tissues, to performing X-ray phase tomography etc. Among those approaches, the phase-stepping technique is commonly used. However, the requirement for multiple image acquisitions and grating scanning introduces mechanical complexity to the system, increases the radiation dose delivered to the samples, and also increases the time for image acquisition. Avoiding phase stepping would significantly increase the measurement speed, especially in X-ray phase tomography [15,16]. Therefore, an important challenge is to find an effective and low-dose approach for phase retrieval in X-ray grating interferometry.

In this Letter, we present a single-shot approach for phase retrieval based on the phase-attenuation duality in high-energy X-ray grating interferometry. The single-shot phase retrieval approach would allow radiation dose reduction and facilitated image acquisition. The essential theoretical framework and the solution for phase retrieval are derived and presented. The feasibility of the proposed approach is then demonstrated by numerically simulating a high-energy X-ray grating interferometer using a photon counting detector.

In X-ray grating interferometry, the measured intensity by each detector pixel can be expressed as [5,16]

$$I_D = \frac{I_0}{M^2}\exp\left(-\int \mu dz\right)\sum_n a_n \cos\left[\frac{2\pi n}{p_2}(x_g + d_T \alpha)\right]$$
$$= \frac{I_0}{M^2}\exp\left(-\int \mu dz\right)\cdot S(x_g + d_T \alpha) \quad (1)$$

where $I_0(x,y)$ is the incident intensity on the object, $M$ is the magnification factor for the object image, $\mu(x,y,z)$ is the linear attenuation coefficient, $x_g$ is the relative transverse position of the two gratings, $d_T$ is the fractional Talbot distance, $p_2$ is the grating period, and the refraction angle $\alpha(x,y)$ is given by the first-order spatial derivative of the phase shift $\Phi(x,y)$ induced by the object [20]

$$\alpha(x,y) = \frac{\lambda}{2\pi M}\frac{\partial \Phi(x,y)}{\partial x} \quad (2)$$

where $\lambda$ is the X-ray wavelength. Note that the spatial dependence has been omitted for the sake of clarity in

Eq.(1). The logarithm of the measured intensity is given by

$$-\ln\left(M^2 I_D / I_0\right) = \int \mu dz - \ln S\left(x_g + d_T \alpha\right) \quad (3)$$

For further analysis, we set the working point $x_g$ at the half-slope position [16], indicated by black dots in Fig. 1(b). In the hypothesis that $\Delta\theta \ll p_2/4d_T$, we can apply the first-order Taylor expansion around $\Delta\theta = 0$ [16], and yield,

$$\ln S\left(x_g + \alpha\Delta\theta\right) \approx \underbrace{\ln S\left(x_g\right)}_{K_1} + \underbrace{\frac{\dot{S}\left(x_g\right)}{S\left(x_g\right)} d_T}_{K_2} \alpha \quad (4)$$

On substitution of Eqs.(2) and (4) into Eq.(3), we obtain

$$K_1 - \ln\left(M^2 I_D / I_0\right) = \int \mu dz - K_2 \frac{\lambda}{2\pi M} \frac{\partial \Phi(x,y)}{\partial x} \quad (5)$$

Furthermore, it is observed that soft tissues encountered in clinical imaging mainly consist of light elements with the atomic number $Z < 10$, such as breast tissue and brain gray/white matter [21]. For those soft tissues, their attenuation cross sections are very well approximated by that of X-ray Compton scattering for X-rays of about 60 keV to 500 keV [22]. Under this circumstance, both the X-ray attenuation and phase shift by soft tissues are all determined by the projected electron density for these high-energy X-rays [22, 23]. When the phase-attenuation duality holds, the X-ray attenuation by the object can be expressed as follows,

$$\int \mu(x,y,z) dz \cong \sigma_{KN} \int \rho_e(x,y,z) dz \quad (6)$$

where $\sigma_{KN}$ is the total cross section for X-ray Compton scattering from a single free electron derived from the Klein-Nishina function [24], and $\rho_e$ is the electron density. In the absence of absorption edges of the materials constituting the object, the X-ray phase shift is linearly proportional to the projected electron density,

$$\Phi(x,y) = \lambda r_e \int \rho_e(x,y,z) dz \quad (7)$$

where $r_e$ is the classical electron radius. By substituting Eqs. (6) and (7) into Eq. (5), we yield the following expression,

$$K_1 - \ln\left(\frac{M^2 I_D}{I_0}\right) = \left(\sigma_{KN} - K_2 \frac{\lambda^2 r_e}{2\pi M} \frac{\partial}{\partial x}\right) \int \rho_e(x,y,z) dz \quad (8)$$

which contains only one unknown, i.e., the projected electron density. Therefore, a single intensity measurement provides a unique solution for the unknown.

From the solution, the X-ray attenuation and phase shift can be retrieved according to Eqs.(6) and (7), respectively.

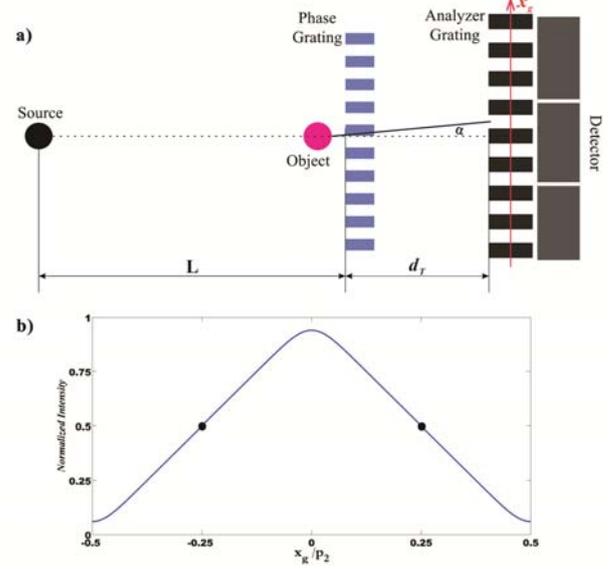

Fig. 1. (Color online) (a) Schematic setup of X-ray grating interferometer with a microfocus source. (b) Normalized intensity as a function of the relative grating position $x_g$.

Consider the one-dimensional (1D) Fourier transform of Eq.(8), and we yield,

$$\hat{\mathbf{F}}\left[K_1 - \ln\left(\frac{M^2 I_D}{I_0}\right)\right] = \left[\sigma_{KN} + i\frac{K_2 \lambda^2 r_e u}{M}\right]\hat{\mathbf{F}}\left(\int \rho_e dz\right) \quad (9)$$

where $\hat{\mathbf{F}}$ denotes the 1D Fourier transform, and $u$ is the spatial frequency. From Eq.(9), we can obtain the Fourier transform of the projected electron density,

$$\hat{\mathbf{F}}\left(\int \rho_e dz\right) = \hat{\mathbf{F}}\left[K_1 - \ln\left(\frac{M^2 I_D}{I_0}\right)\right] \bigg/ \left(\sigma_{KN} + iK_2 \lambda^2 r_e u/M\right) \quad (10)$$

And then the X-ray phase shift can be retrieved by

$$\Phi = \hat{\mathbf{F}}^{-1}\left\{\hat{\mathbf{F}}\left[K_1 - \ln\left(\frac{M^2 I_D}{I_0}\right)\right] \bigg/ \left(\frac{\sigma_{KN}}{\lambda r_e} + \frac{iK_2 \lambda u}{M}\right)\right\} \quad (11)$$

with $\hat{\mathbf{F}}^{-1}$ denoting the 1D inverse Fourier transform.

To test the feasibility of the proposed phase retrieval approach, we performed numerical experiments. The simulated grating interferometer consists of a microfocus source with a 10 μm focal-spot size, a π/2-shifting phase grating, an ideal absorption grating [PRA2015] and a photon counting detector with a 50 μm pixel size. To obtain a 60 keV design energy, we modeled a tungsten target/anode X-ray source operating at 120 kVp [25]. A 4 mm aluminum (Al) filter was utilized to block the low energy photons and harden the X-ray beam. The beam hardening due to the silicon substrate of the two gratings was also taken into account. The detective quantum efficiency of the detector is approximated by the photon absorption probability in a 750-µm-thick cadmium-zinc-telluride (CZT) sensor. Photon shot noise, which is signal-dependent and Poisson distributed, has been added to the simulated intensity. A test phantom consists of two cylindrical rods-breast tissue, brain grey/white matter-with a 2.0 mm diameter, is constructed following the reference [2, 21]. The material compositions, energy-dependent attenuation coefficients and refractive index decrements are calculated by using the data provided by the NIST [26].

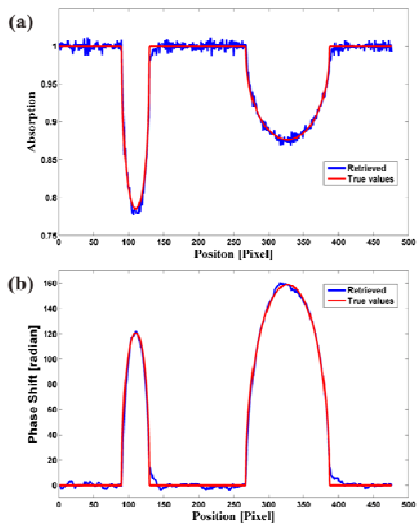

Fig. 2. Profiles of the retrieved values with the proposed approach and the true values of (a) X-ray absorption and (b) phase shift at 60 keV.

Figure 2 shows the profiles of the retrieved X-ray absorption and phase shift of the test object together with the corresponding theoretical true values, by plotting the pixel value as a function of the position. A good agreement is achieved for both absorption and phase shift of the considered materials, confirming the feasibility of the proposed approach. Furthermore, the relative errors of the retrieved values with respect to the theoretical true values are calculated, to demonstrate the quantitativeness of the proposed approach. Table 1 summarizes the calculated errors for each material by averaging the relative errors over the associated line profiles. The errors of the retrieved phase shift are relatively greater than those of the retrieved absorption. As seen from Fig. 2, the relative errors of the retrieved values are particularly large at the edge of the cylinders. This is due to the low-pass filtering property of the inverse differential operator, and might be overcome by using detectors with a smaller pixel size.

Table 1. Calculated relative errors of retrieved absorption and phase shift of the phantom with respect to the true values.

| Materials | Relative Errors | |
|---|---|---|
| | Absorption | Phase shift |
| Breast Tissue | 0.45% | 5.94% |
| Brain Gray/White Matter | 0.07% | 1.84% |

In summary, we presented a single-shot approach for quantitative phase retrieval in high-energy X-ray grating interferometry. The feasibility of this approach using phase-attenuation duality is verified by numerical experiments for different materials. Owing to its advantages of simplicity and applicability to a wide range of material compositions, we can expect widespread potential applications of this method in biomedical imaging, materials science, etc. Furthermore, the proposed approach can be straightforwardly generalized to two-dimensional X-ray grating interferometry [27,28], and edge illumination method [29], and extended to tomographic imaging of the related physical quantities.

**Funding.** National Natural Science Foundation of China (11475170, U1532113, 11205157); Anhui Provincial Natural Science Foundation (1508085MA20).